\documentclass[doublecol]{epl2}
\usepackage{graphics,epsfig}
\usepackage{dcolumn}
\usepackage{bm}
\usepackage{color}
\usepackage{subfig}
\usepackage{CJK}
\newcommand{\PreserveBackslash}[1]{\let\temp=\\#1\let\\=\temp}
\newcolumntype{C}[1]{>{\PreserveBackslash\centering}p{#1}}
\newcolumntype{R}[1]{>{\PreserveBackslash\raggedleft}p{#1}}
\newcolumntype{L}[1]{>{\PreserveBackslash\raggedright}p{#1}}

\title{Optimal forwarding ratio on dynamical networks with heterogeneous mobility}
\shorttitle{Optimal forwarding ratio on heterogeneous dynamical networks}
\author
{
    Yu Gan\inst{1} \and
    Ming Tang\inst{1, 2}\footnote{\email{E-mail:tangminghuang521@hotmail.com}} \and
    Han-Xin Yang\inst{3}
}
\shortauthor{Yu Gan \etal}

\institute
{
  \inst{1} Web Sciences Center, University of Electronic Science and Technology of China - Chengdu 610051, P. R. China\\
  \inst{2} Department of Physics and Institute of Theoretical Physics, The Chinese University of Hong Kong, Shatin, Hong Kong, P. R. China\\
  \inst{3} Department of Physics, Fuzhou University - Fuzhou 350002, P. R. China
}


\abstract{As the discovery of non-Poissonian statistics of human mobility trajectories, more attention has been paid to understanding the role of these patterns in different dynamics. In this study, we first introduce the heterogeneous mobility of mobile agents into dynamical networks, and then investigate the forwarding strategy on the heterogeneous dynamical networks. We find that the faster speed and the higher proportion of high-speed agents can enhance the network throughput and reduce the mean traveling time in the case of random forwarding. A hierarchical structure in the dependence of high-speed is observed: the network throughput remains unchanged in small and large high-speed value. It is interesting to find that the slightly preferential forwarding to high-speed agents can maximize the network capacity. Through theoretical analysis and numerical simulations, we show that the optimal forwarding ratio stems from local structural heterogeneity of low-speed agents.}

\begin{document}
\begin{CJK}{GBK}{song}

\maketitle{}

\section{Introduction}
The notion of mobile agent is really important in the study of many dynamical systems, such as collective motion in biology~\cite{Vicsek2012}, information transmission on wireless ad-hoc network~\cite{Abolhasan2004}, and traffic congestion in city transportation system~\cite{Barthelemy2011}. In all of these dynamical systems, each mobile agent is free to move independently in any direction, and locally interacts with current nearby agents. Take wireless ad-hoc network for example~\cite{Abolhasan2004}. The continuous movement of mobile devices would frequently change their links to other devices due to limited forwarding range. Thus, these dynamical systems can be treated primitively as dynamical networks, in which connections between nodes always change temporarily. Network theory provides therefore a natural framework to study the emergence of structural and dynamical properties of these systems~\cite{Albert2002,boccaletti2006,dorogovtsev2008,Barrat2008,Newman2010}.

Physicists recently became increasingly fascinated by the dynamics on dynamical networks, such as epidemic spreading of mobile individuals~\cite{Frasca2006,Buscarino2008,Peruani2008,Li2010,Yang2012}, synchronization of mobile oscillators~\cite{Frasca2008,Fujiwara2011,Frasca2012}, and consensus of communicating agents~\cite{Baronchelli2012}. Especially for packet forwarding of mobile devices, how to enhance transmission efficiency has become a crucial problem due to the rapid development of wireless communication technology. Considering costs and technology limitations, more efforts should be made to improve the routing strategy~\cite{Abolhasan2004,Echenique2004,Echenique2005,Yan2006,Huan2007,Tang2009,Tang2011}. Yang \emph{et al.} first studied transportation dynamics on dynamical networks~\cite{Yang2011}, and compared random routing with greedy routing~\cite{Yang2011-2}.

However, the previous researches have only focused on the case of identical mobile agents, which can not reflect the diversity of mobile individuals. Since the discovery of non-Poissonian statistics of human behaviors such as human interaction activities~\cite{Barabasi2005} and mobility trajectories~\cite{Brockmann2006,Gonzalez2008}, more and more scientists have been paying attention to the role of these patterns in different dynamics. Most recent research results showed that both time and space activities have significant impacts on spreading dynamics~\cite{Vazquez2007,Iribarren2009,Tang2009-1,Tang2009-2,Wang2009,Balcan2011,Belik2011,Zhao2012}. In this letter, we introduce the heterogeneous mobility of mobile agents into dynamical networks, and then propose an optimal forwarding strategy on the dynamical networks with heterogeneous mobility.

\section{Model}
We consider $N$ agents performing random walk in an area of $L
\times L$ with periodic boundary conditions.  Initially, the
agents are randomly put on the area.
Let $v_i(t)$ be the speed and $\theta_i(t)$ be the angle with
respect to the $x$-axis characterizing the direction of the
velocity of agent $i$, and $x_{i}(t)$ and $y_{i}(t)$ be the
coordinates of the agent $i$ and time $t$.  The coordinates and
the velocity of agent $i$ at time $t+1$ are updated according to
\begin{eqnarray}\label{eq:position}
x_i(t+1)&=&x_i(t)+v_i(t)cos\theta_i(t), \nonumber \\
y_i(t+1)&=&y_i(t)+v_i(t)sin\theta_i(t), \nonumber \\
v_i(t+1)&=&v_i, \nonumber \\
\theta_i(t+1)&=&\xi_i(t),
\end{eqnarray}
where $\xi_i(t)$ is a random variable chosen uniformly between
the interval  $[-\pi,\pi]$ after the coordinates are updated.
To account for heterogeneous mobilities of the mobile agents
~\cite{Brockmann2006,Gonzalez2008}, we assume that the agents
only take on either a low speed $v_l$ or a high speed $v_h$.  The
fraction of high speed agents is $f$, i.e., the system is
characterized by a distribution of speeds $D(v)$ among the agents
represented by
\begin{equation}\label{eq:speed}
D(v) = (1-f) \delta(v - v_l) + f \delta(v-v_h),
\end{equation}
where $\delta(x)$ is the Dirac $\delta$-function.  The speed of
an agent, once assigned initially, does not vary with time.
Therefore, an agent moves with a constant speed but in random
directions.

To consider the packet routing efficiency in a network of mobile
agents, let $R$ packets be generated at a time step.  Each of
these packets is generated randomly at some agent $i$ and placed
at the end of the queue of agent $i$, with a destination that is
also randomly chosen among the agents in the system.  To forward
the packet to its destination, each agent is capable of
forwarding at most $C$ packets in its queue to selected neighbors
every time step.  The delivery follows a First-In-First-Out
policy.  As the agents are mobile and in view of the connectivity
of electronic devices, we regard the agents that are
instantaneously within an area of radius $r$ of an agent $i$ to
be the neighbors, i.e., the neighbors of agent $i$ are those with
$d_{ij} < r$, where $d_{ij}$ is the distance between agent $i$
and agent $j$.  The radius $r$ thus sets the range over which a
packet could be forwarded in a time step. In this point of view,
a mobile-agent network corresponds to a dynamical network, with
the links constantly broken and established as the agents move.
The algorithm for forwarding a packet goes as follows. If the
destination of a packet is found among the neighbors, the packet
will be delivered to its destination and removed immediately.  If
not, an agent decides whether to forward it to a high-speed
neighbor with a probability $p$ or to a low-speed neighbor with a
probability $1-p$.  After making a decision, a neighbor of the
chosen type is randomly picked and then the forwarded packet is
put at the end of the queue of the selected neighbor. If there is
no neighbor of the chosen type, the packet will not be delivered.
The process is repeated for each packet to be forwarded and every
agent carries out the algorithm in every time step.  As there is a
fraction $f$ of high-speed agents in the system, the case of $p =
f$ roughly equal to the case of randomly choosing a neighbor to
forward a packet regardless of its type, and the cases of $p<f$
($p>f$) correspond roughly to forwarding a packet to low-speed
(high-speed) neighbors preferentially.

Let $S(t)$ be the total number of packets in the system at time
$t$.  In the congested phase, there will be more and more packets
accumulated in the system; while in the balanced phase, the
number of delivered and removed packets balances that of packets
generated.  These two phases can be characterized by
~\cite{Tang2011}
\begin{equation}\label{eq:order parameter}
\eta = \frac{C}{R}\lim_{t\rightarrow \infty} \frac{\langle
S(t+\Delta t) - S(t)\rangle}{\Delta t},
\end{equation}
which plays the role of the order parameter in that $\eta >0$ in
the congested phase and $\eta = 0$ in the balanced phase.  For a
given value of $C$, the system is in the balanced (congested)
phase for packet generation rates $R < R_{c}$ ($R > R_{c}$).  A s
a figure of merit for packet delivery, a higher $R_{c}$
corresponds to a better algorithm.

\section{Simulation Results}
First we investigate the high-speed agents how to influence network throughput $R_c$ in the case of random forwarding, i.e., $p=f$. Fig. 1(a) depicts the phase transition (from free flow state to congestion state) for different $v_h$ values when $f=0.4$, and $R_c$ is found to depend on $v_h$. A hierarchical structure in the dependence is shown in Fig. 1(b): $R_c$ remains unchanged in small or large $v_h$ value, which is consistent with the results in Ref.~\cite{Yang2011}. Besides, Figs. 2 (a) and (b) show that $R_c$ increases with $f$, which denotes more high-speed agents can enhance the network throughput. Then the question is: Why does the speed $v_h$ and the proportion $f$ of high-speed agents influence the network throughput? We will discuss it later (see Eq.~(\ref{eq:throughput2})).

\begin{figure}
\begin{center}
\scalebox{0.38}[0.39]{\includegraphics{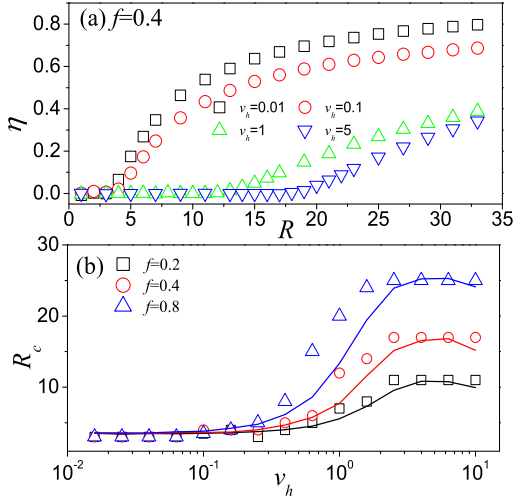}} \caption{(Color
online) The network throughput as a function of high speed value for the case of random forwarding. (a) Order parameter $\eta(R)$ versus $R$ for different $v_h$ values when $f=0.4$. $\eta(R)$ is obtained by averaging over $10^3$ time steps after disregarding $3.5\times10^3$ initial steps as transients. (b) The network throughput $R_c$ versus $v_h$ when $f=0.2, 0.4, 0.8$. The lines are the theoretical predictions from Eq.~(\ref{eq:relationship}) where $\langle T\rangle$, $k_{l,max}^{e}$ and $k_{h,max}^{e}$ are obtained by numerical simulations. The parameters are chosen as $N=10^3, L=10, r=1, v_l=0.001, C=1, p=f$. The results are obtained by averaging over $10$ independent realizations.} \label{fig1}
\end{center}
\end{figure}

\begin{figure}
\begin{center}
\scalebox{0.38}[0.39]{\includegraphics{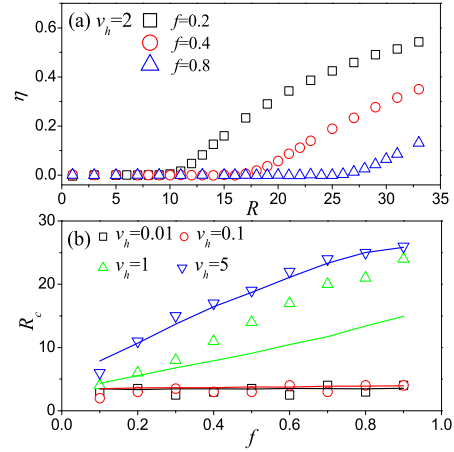}} \caption{(Color
online) The network throughput as a function of the proportion of high-speed agents for the case of random forwarding. (a) Order parameter $\eta(R)$ versus $R$ for different $f$ values when $v_h=2$. $\eta(R)$ is obtained by averaging over $10^3$ time steps after disregarding $3.5\times10^3$ initial steps as transients. (b) The network throughput $R_c$ versus $f$ when $v_h=0.01, 0.1, 1, 5$. The lines are the theoretical predictions from Eq.~(\ref{eq:relationship}) where $\langle T\rangle$, $k_{l,max}^{e}$ and $k_{h,max}^{e}$ are obtained by numerical simulations. The parameters are chosen as $N=10^3, L=10, r=1, v_l=0.001, C=1, p=f$. The results are obtained by averaging over $10$ independent realizations.} \label{fig2}
\end{center}
\end{figure}

Next, we think of the case of preferential forwarding, i.e., $p\neq f$. As shown in Fig. 3(a), it is very interesting to note that there is a maximum $R_c$ value when $p$ is slightly greater than $f$, e.g., $R_m\approx15$ at $p_m\approx0.48$ when $f=0.4$ and $R_m\approx22$ at $p_m\approx0.88$ when $f=0.8$. As $p_m>f$ corresponds to the preferential forwarding to high-speed agents in Fig. 3(b), this optimal phenomenon means that the slightly preferential forwarding can enhance network throughput to some extent. For example, $R_m\approx22 (p\approx0.88)>R_c\approx16 (p=0.8)$ when $f=0.8$. Moreover, Fig.~\ref{fig4} shows that faster speed and the higher proportion of high-speed agents can reduce the mean traveling time $\langle T\rangle$ (defined as the average number of hops for all data packets from their sources to destinations) in the case of random forwarding.

\section{Theory}
In order to understand the above optimal phenomenon, here we present a mean field analysis of network throughput. As all agents are randomly distributed on the planar space at any time $t$ due to the random walk of agents, the degree distribution will be approximated by~\cite{Antonioni2012}
\begin{equation}\label{eq:degree distribution}
P(k) = \frac{e^{-\langle k\rangle}{\langle k\rangle}^k}{k!},
\end{equation}
where $\langle k\rangle=N\pi r^2/L^2$.

\begin{figure}
\begin{center}
\scalebox{0.38}[0.39]{\includegraphics{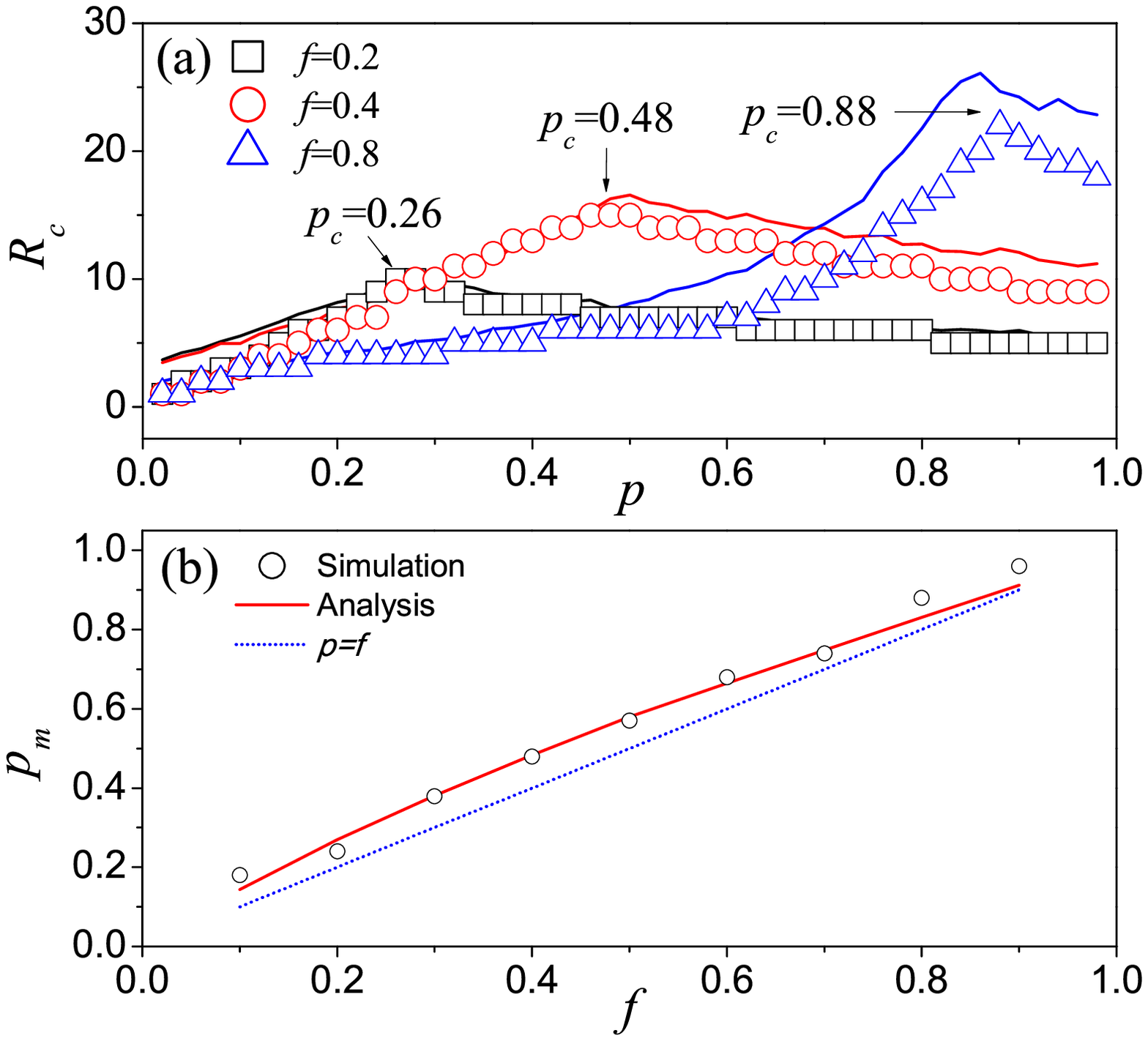}} \caption{(Color
online) The network throughput as a function of the forwarding rate when $f$ is fixed. (a) The network throughput $R_c$ versus $p$ when $f=0.2, 0.4, 0.8$. The lines are the theoretical predictions from Eq.~(\ref{eq:throughput2}) where $\langle T\rangle(C=\infty)$ is obtained by numerical simulations (see Fig.~\ref{fig3}) and $k_{l,max}^{e}$ is given by Eq.~(\ref{eq:kmax}). A difference between the theoretical predictions and the numerical results is originated from the real $\langle T\rangle_{real}>\langle T\rangle(C=\infty)$ at the critical point. (b) The optimal forwarding $p_m$ versus $f$. The red solid line is the theoretical prediction from Eq.~(\ref{eq:optimal}) where $k_{l,max}^{e}$ is given by Eq.~(\ref{eq:kmax}), and the blue dotted line corresponds to $p=f$. The parameters are chosen as $N=10^3, L=10, r=1, v_l=0.001, v_h=2, C=1$. The results are obtained by averaging over $10$ independent realizations.} \label{fig3}
\end{center}
\end{figure}

\begin{figure}
\begin{center}
\scalebox{0.38}[0.39]{\includegraphics{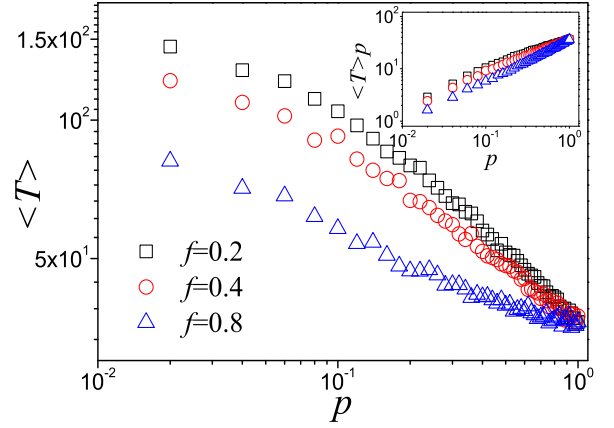}} \caption{(Color
online) The mean traveling time $\langle T\rangle$ as a function of the forwarding rate $p$ in the free flow state with $C=\infty$ when $f=0.2, 0.4, 0.8$. The inset shows $\langle T\rangle p$ as a function of the forwarding rate $p$. The parameters are chosen as $N=10^3, L=10, r=1, v_l=0.001, v_h=2$. The results are obtained by averaging over $10$ independent realizations.} \label{fig4}
\end{center}
\end{figure}

In the balance phase, according to the forwarding strategy, the evolution of the packet numbers of low-speed agent $l$ with $k_l(t)$ and high-speed agent $h$ with $k_h(t)$ can be written as
\begin{eqnarray}\label{eq:rate equation}
\frac{dn_l(t)}{dt}&=&-n_l(t)+\sum_{j=1}^{N}\frac{(1-p)A_{lj}n_j(t)}{(1-f)k_j(t)},\nonumber \\
\frac{dn_h(t)}{dt}&=&-n_h(t)+\sum_{j=1}^{N}\frac{p A_{hj}n_j(t)}{fk_j(t)},
\end{eqnarray}
where the sum runs over all nodes of the network, and $A_{lj}$ is the element of the adjacency matrix, in which $A_{lj}=1$ when $d_{lj}<r$, otherwise $A_{lj}=0$. At large time $t$, all agents will reach the balance of traffic flow, and we have
\begin{eqnarray}\label{eq:stationary state}
n_l(t)&=&\sum_{j=1}^{N}\frac{(1-p)A_{lj}n_j(t)}{(1-f)k_j(t)},\nonumber \\
n_h(t)&=&\sum_{j=1}^{N}\frac{p A_{hj}n_j(t)}{fk_j(t)}.
\end{eqnarray}
Considering $Nf$ high-speed agents randomly distributed on the planar space, we know that an agent with $k(t)$ has $fk(t)$ high-speed neighbors and $(1-f)k(t)$ low-speed neighbors. We suppose $n_l(t)=B_1k_l(t)$ and $n_h(t)=B_2k_h(t)$, and then Eq.~(\ref{eq:stationary state}) is transformed into
\begin{eqnarray}\label{eq:solution}
n_l(t)&=&\frac{(1-p)k_l(t)}{1-f}[(1-f)B_1+fB_2],\nonumber \\
n_h(t)&=&\frac{pk_h(t)}{f}[(1-f)B_1+fB_2],
\end{eqnarray}
where the occupation packet number is independent of degree correlations~\cite{Antonioni2012}. From the above, we know that $B_1/B_2=[(1-p)f]/[p(1-f)]$, and Eq.~(\ref{eq:solution}) is also written as
\begin{eqnarray}\label{eq:packet distribution}
n_l(t)&=&\frac{B(1-p)k_l(t)}{1-f},\nonumber \\
n_h(t)&=&\frac{Bpk_h(t)}{f},
\end{eqnarray}
where $B$ is a constant. In the balance phase, there are $S=\sum_{l}n_l(t)+\sum_{h}n_h(t)\simeq R\langle T\rangle$ packets on the network at each time step~\cite{Zhao2005}. Submitting Eqs.~(\ref{eq:degree distribution}) and (\ref{eq:packet distribution}) into this self-consistent relationship, and we obtain
\begin{equation}\label{eq:B}
B=\frac{R\langle T\rangle}{N\langle k\rangle}.
\end{equation}

At the critical point, the dynamical balance of traffic flow happens in a relatively long time with the same order as $O(\langle T\rangle)$. The node possessing the maximum occupation packet number must meet the relationship ${\sum_t}^{O(\langle T\rangle)} n_{max}(t)\leq CO(\langle T\rangle)$ in $O(\langle T\rangle)$ steps. According to Eqs.~(\ref{eq:packet distribution}) and (\ref{eq:B}), we have
\begin{eqnarray}\label{eq:relationship}
\frac{R\langle T\rangle pk_{h,max}^{e}}{N\langle k\rangle f}\leq C,\nonumber \\
\frac{R\langle T\rangle(1-p)k_{l,max}^{e}}{N\langle k\rangle(1-f)}\leq C,
\end{eqnarray}
where $k_{h,max}^{e}$ and $k_{l,max}^{e}$ are the maximum effective degree of high-speed agents and low-speed agents respectively, and the effective degree of node $i$ is defined as $k_i^e=\frac{1}{O(\langle T\rangle)}\sum_{t}k_i(t)$.

Owing to the heterogeneous mobility, $k_{l,max}^{e}$ of low-speed agents is different from that of high-speed agents. All high-speed agents can span a broad area of planar space in $O(\langle T\rangle)$ steps due to their fast moving. According to the law of large number, we have
\begin{equation}\label{eq:high distribution}
k_h^e=\lim_{O(\langle T\rangle)\rightarrow\infty} \frac{1}{O(\langle T\rangle)}\sum_{t}k_h(t)\simeq\langle k\rangle.
\end{equation}
The maximum effective degree of high-speed agents $k_{h,max}^e$ is approximatively equal to $\langle k\rangle$. On the other hand, low-speed agents are taken as motionless nodes in this stage because of its very small speed. For this reason, the effective degree of a low-speed agent is divided into two parts: $f\langle k\rangle$ high-speed neighbors and $k'_l$ settled low-speed neighbors. The number of settled low-speed neighbors follows a Poisson distribution
\begin{equation}\label{eq:low distribution}
P(k'_l) = \frac{e^{-\langle k'_l\rangle}{\langle k'_l\rangle}^{k'_l}}{k'_l!},
\end{equation}
where $\langle k'_l\rangle=(1-f)N\pi r^2/L^2$. Thus, the distribution of $k_l^e$ follows
\begin{equation}\label{eq:low-effective-distribution}
P(k_l^e) = \frac{e^{-\langle k'_l\rangle}{\langle k'_l\rangle}^{(k_l^e-f\langle k\rangle)}}{(k_l^e-f\langle k\rangle)!}.
\end{equation}
The natural cut-off effective degree of low-speed agents can be calculated by~\cite{Boguna2004}
\begin{equation}\label{eq:kmax}
N(1-f)\int_{k_{l,max}^{e}}^{\infty}P(k_{l}^{e})dk_{l}^{e}\sim1.
\end{equation}

Submitting Eq.~(\ref{eq:high distribution}) into Eq.~(\ref{eq:relationship}), the network throughput is given by
\begin{equation}\label{eq:throughput1}
R_c=Min\{R_c^h,R_c^l\},
\end{equation}
and
\begin{eqnarray}\label{eq:throughput2}
R_c^h&=&\frac{NCf}{\langle T\rangle p},\nonumber \\
R_c^l&=&\frac{N\langle k\rangle C(1-f)}{\langle T\rangle(1-p)k_{l,max}^{e}},
\end{eqnarray}
where $Min\{\}$ represents the less of the two. If $p=f$, Eq.~(\ref{eq:throughput2}) will go back to the case of random forwarding. When $v_h$ is much smaller than $r$, the dynamic network is approximatively taken as a fixed structure in a comparatively long time, so the network throughput changes very little. As $v_h$ increase to the value with the same order as $r$, the shorter mean traveling time, which is due to the longer forwarding distance in each hop, results in the increase of network throughput. For large values of $v_h$, the network throughput remains unchanged because of the fixed mean traveling time and maximum degree~\cite{Yang2011}. For this reason, a hierarchical structure in the dependence of high-speed is observed in Fig.~\ref{fig1}(b). In this case, $\langle T\rangle$ (due to more high-speed agents) and $k_{l,max}^{e}$ (from Eq.~(\ref{eq:kmax})) decrease with $f$. Thus, $R_c=R_c^l$ increases with $f$, which is validated by the results in Fig.~\ref{fig2}(b). From Eq.~(\ref{eq:relationship}), the theoretical predictions are basically consistent with the numerical results in Figs.~\ref{fig1}(b) and \ref{fig2}(b). When $v_h$ is not very large (e.g., $v_h=1$), there is a difference between the theoretical predictions and the numerical results due to the invalidation of effective degree hypothesis.

When $f$ is fixed, $\langle T\rangle$ decreases with $p$ because the preferential forwarding to high-speed agents can increase the distance which packets are forwarded in one time step (see numerical results in Fig.~\ref{fig4}). From Eq.~(\ref{eq:throughput2}), we know that $R_c^h$ decreases with $p$ because of the increase of $\langle T\rangle p$ (confirmed by numerical results in the inset of Fig.~\ref{fig4}), while $R_c^l$ increases with $p$ because of the decrease of $\langle T\rangle(1-p)$. There exits a maximum network throughput $R_m$ when $R_c^l=R_c^h$ (see Fig.~\ref{fig3} (a)), and the corresponding optimal $p_m$ is thus obtained by
\begin{equation}\label{eq:optimal}
p_m=\frac{fk_{l,max}^{e}}{(1-f)\langle k\rangle+fk_{l,max}^{e}}.
\end{equation}
Interestingly, $p_m$ only depends on the structure of the dynamical network.
Owing to the Poisson characteristic of effective degree of low-speed agents, i.e., $k_{l,max}^{e}>\langle k\rangle$, $p_m$ will be slightly greater than $f$, which means that the optimal forwarding rate is the slightly preferential forwarding to high-speed agents. As shown in Fig.~\ref{fig3} (b), the theoretical predictions are in good agreement with the simulation results.

\begin{figure}
\begin{center}
\scalebox{0.38}[0.39]{\includegraphics{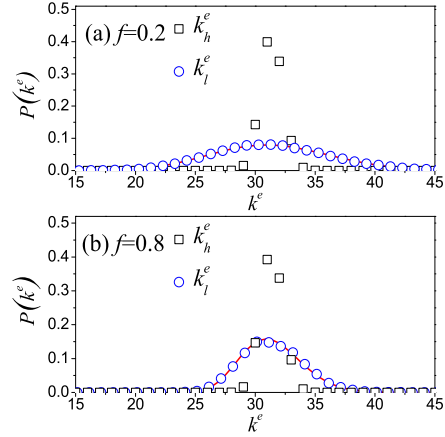}} \caption{(Color
online) Effective degree distribution for two kinds of agents when $f=0.2$ (a) and $f=0.8$ (b). The parameters are chosen as $N=10^3, L=10, r=1, v_l=0.001, v_h=2, O(\langle T\rangle)=50$. The results are obtained by averaging over $10^2$ independent realizations. The lines are the theoretical predictions from Eq.~(\ref{eq:low-effective-distribution}).} \label{fig5}
\end{center}
\end{figure}

\section{Conclusion and Discussion}
In conclusions, we have investigated the forwarding strategy on the dynamical networks with heterogeneous mobility. First, we have shown the faster speed and the higher proportion of high-speed agents can enhance the network throughput and reduce the mean traveling time in the case of random forwarding. A hierarchical structure in the dependence of high-speed is observed: the network throughput remains unchanged in small and large high-speed value. Second, it would be interesting to know that the slightly preferential forwarding to high-speed agents can maximize the network capacity. By combining theoretical analysis with numerical simulations, we have obtained the optimal forwarding rate, and found this optimal phenomenon stems from local structural heterogeneity of low-speed agents. This work provides us further understanding and new perspective in the effect of human dynamics on the forwarding strategy on dynamical networks, and thus may help to optimize the delivering strategy on heterogeneous dynamical networks.

\acknowledgments
Ming Tang would like to thank Pakming Hui for stimulating discussions. This work is supported by the NNSF of China (Grants No. 11105025),
China Postdoctoral Science Foundation (Grant No. 20110491705),
the Specialized Research Fund for the Doctoral Program of Higher Education
(Grant No. 20110185120021), China Postdoctoral Science Special Foundation
(Grant No. 2012T50711), and the Fundamental Research Funds for the Central Universities
(Grant No. ZYGX2011J056).

\end{CJK}
\end{document}